\documentstyle[prd,aps,multicol,tighten,epsf,rotate]{revtex}

\begin{document}
\draft

%
\title{Cosmic string induced sheet like baryon inhomogeneities at
quark-hadron transition}
\author{Biswanath Layek \footnote{e-mail: layek@iopb.res.in},
Soma Sanyal \footnote{e-mail: sanyal@iopb.res.in},
and Ajit M. Srivastava \footnote{e-mail: ajit@iopb.res.in}}
%
\address{Institute of Physics, Sachivalaya Marg, Bhubaneswar 751005, 
India}
%
%
\maketitle
\widetext
\parshape=1 0.75in 5.5in
\begin{abstract}
Cosmic strings moving through matter produce wakes where density is higher 
than the background density. We investigate the effects of such wakes 
occurring at the time of a first order quark-hadron transition in the 
early universe and show that they can lead to separation of 
quark-gluon plasma phase in the wake region, while the region outside
the wake converts to the hadronic phase. Moving interfaces
then trap large baryon densities in sheet like regions which
can extend across the entire horizon. Typical separation between 
such sheets, at formation, is of the order of a km. Regions of baryon 
inhomogeneity of this nature, i.e. having a planar geometry, and separated 
by such large distance scales, appear to be well suited for the recent 
models of inhomogeneous nucleosynthesis to reconcile with the large 
baryon to photon ratio implied by the recent measurements of the cosmic 
microwave background power spectrum.

\end{abstract}
\vskip 0.125 in
\parshape=1 0.75in 5.5in
\begin{multicols}{2}
\narrowtext

\section{Introduction}

 Generation of baryon inhomogeneities during quark-hadron transition is an 
important issue due to its consequences for the nucleosynthesis, and the 
possibility of creating compact baryon rich objects \cite{wtn}. Recently, 
there has been a renewed interest in the investigation of nucleosynthesis
in the presence of baryon number inhomogeneities, the so called inhomogeneous
big bang nucleosynthesis (IBBN) \cite{ibbn1,ibbn2}. This originates from 
the recent measurements of the angular power spectrum of the cosmic microwave
background radiation (CMBR) by the BOOMERANG, and MAXIMA experiments
\cite{expt}, which lead to a value of baryon density to critical density 
ratio $\Omega_b$ which is larger than the value allowed by the 
conventional models of homogeneous nucleosynthesis.
Earlier, there have been numerous investigations of the nature of baryon 
inhomogeneities generated during a first order quark-hadron phase transition  
\cite{bfluct} and its effects on the abundances of various elements 
\cite{nsynth}. In these investigations, baryon inhomogeneities 
arise due to moving bubble walls at the transition, with baryons getting 
concentrated in the remaining localized quark-gluon plasma (QGP) regions. 
Typical separation, $l_{inhm}$, between such localized baryonic lumps is 
of the order of separation between the nucleation sites of the hadronic
bubbles, which is at most of the order of few cm at the end of the 
quark-hadron transition for homogeneous nucleation\cite{bfluct,impur}. 
In order that these baryonic lumps survive various dissipative processes, 
such as baryon diffusion, neutrino inflation etc., up to the time of 
nucleosynthesis, the required separation of these lumps is of the order 
of few meters \cite{bfluct}. 

  There have been some attempts of getting a larger separation between
baryon inhomogeneity regions, $l_{inhm}$, by deviating form the 
homogeneous nucleation theory. Christiansen and Madsen have 
argued \cite{impur} that sufficiently large values of $l_{inhm}$ can arise
in the presence of impurities due to heterogeneous nucleation of hadronic
bubbles. It is mentioned in ref.\cite{impur} that possible sources of such 
impurities could be primordial black holes, cosmic strings, magnetic
monopoles, or relic fluctuations from the electroweak scale. Hadronic 
bubbles  are expected to nucleate at these impurities with enhanced rates,
leading to separation between baryon rich regions being of the order of
typical separation between impurities. Recently, Ignatius and Schwarz have 
proposed \cite{inhm} that the presence of density fluctuations (which 
could arise at the time of inflation) at  
quark-hadron transition  will lead to splitting of the region in 
hot and cold regions with cold regions converting to hadronic phase 
first. Baryons will then get trapped in the (initially) hotter regions. 
Estimates of sizes and separations of such density fluctuations were made 
in ref.\cite{inhm} using COBE measurements of the temperature fluctuations
in the cosmic microwave background radiation arising from the density
fluctuations responsible for the structure formation in the universe.
It was argued in ref.\cite{inhm} that it is possible to get $l_{inhm}$
as large as few meters.

 There are many possible sources of density fluctuations in the early
universe, inflation being one such possibility. Cosmic defects (strings,
monopoles, textures) also lead to density fluctuations in the universe.
There has been extensive study of density fluctuations generated by cosmic 
strings from the point of view of structure formation \cite{str1}. In this 
paper we study the influence of density fluctuations produced by 
cosmic strings on the dynamics of the quark-hadron transition. Though recent 
measurements of temperature anisotropies in the microwave background 
by BOOMERANG, and MAXIMA experiments \cite{expt} at angular scales
of $\ell \simeq$ 200 disfavor models of structure formation based 
exclusively on cosmic strings \cite{str2,str3}. Still, due to 
many uncertainties in the scaling
models of cosmic string network evolution one can not rule them out
as candidates of sources of required density fluctuations. Some feature
of the evolving string network which is capable of introducing super horizon
string correlations, may be able to make these models compatible with
recent observations. Further, even with present models, it is not ruled 
out that cosmic strings may contribute to some part in the structure 
formation in the universe. Above all, cosmic strings generically arise
in many Grand Unified Theory (GUT) models. If the GUT scale is somewhat
lower than $10^{16}$ GeV then the resulting cosmic strings will not
be relevant for structure formation, but they will still affect various
stages of the evolution of the universe in important ways. (For a 
discussion of these issues, see \cite{str3}.) It is in this spirit 
that we undertake the study of the effects of cosmic strings
on quark-hadron phase transition. 

We will show that very interesting effects can arise due to cosmic strings, 
such as generation of sheet like baryonic inhomogeneities spreading across 
the horizon. Typical separation between such sheets will be of the order of 
a km at quark-hadron transition stage. (For simplicity we do not 
consider here the effects of density fluctuations due to oscillating
string loops. These will give rise to localized regions of baryon
inhomogeneity.)  There is no other model known
which can lead to such large values of separations between baryon 
inhomogeneity regions. Such inhomogeneities will have important effects 
on the nucleosynthesis. One could also reverse the argument and study the
constraints on various cosmic string parameters which will arise from 
various elemental abundances from inhomogeneous nucleosynthesis in 
such a model. 

 As we mentioned above, a re-consideration of the models of
IBBN is well motivated at the present stage due to recent measurements
of the angular power spectrum of CMBR \cite{expt}. These results imply
a value of $\Omega_b$ which is larger than that allowed by the conventional
homogeneous nucleosynthesis. In this context we mention that a recent 
analysis by Kurki-Suonio and Sihvola \cite{ibbn1} shows that the
results of IBBN may be (marginally) compatible with this large value of
$\Omega_b$ as required by the recent CMBR measurements \cite{expt}. 
However, as pointed out in ref.\cite{ibbn1}, this agreement can be
achieved only if regions of baryon inhomogeneity are separated by a 
distance scale of about 35-70 km (at T = 1 MeV), and if these regions
have a more planar structure with high surface to volume ratio.
It is intriguing that the baryon inhomogeneities generated by cosmic
strings in our model satisfy both these criterion. These are clearly
planar structures. Further, distance between these sheets at quark-hadron
transition being of order 1 km (or somewhat smaller) will translate to
about 100 km distance scale at T = 1 MeV.

\section{Cosmic string wakes in a relativistic fluid}

 We first briefly review the structure of density fluctuations 
produced by a cosmic string moving through a relativistic fluid.
The space-time around a straight cosmic string (along 
the z axis) is given by the following metric \cite{mtrc},

\begin{equation}
ds^2 = dt^2 - dz^2 - dr^2 -  (1 - 4G\mu)^2 r^2 d\phi^2 ,
\end{equation}

\noindent where $\mu$ is the string tension. This metric describes a conical 
space, with a deficit angle of $8\pi G\mu $.  This metric can be put in the 
form of the Minkowski metric by defining angle $\phi^\prime = (1-4G\mu) 
\phi$. However, now $\phi^\prime$ varies between 0 and $(1-4G\mu)2\pi$, that 
is, a wedge of opening angle $8\pi G\mu$ is removed from the Minkowski 
space, with the two boundaries of the wedge being identified. It is well 
known that in this space-time, two geodesics going along the opposite sides 
of the string, bend towards each other \cite{gdsk}. This results in binary 
images of distant objects, and can lead to planar density fluctuations. These
wakes arise as the string moves through the background medium, giving
rise to velocity impulse for the particles in the direction of the 
surface swept by the moving string. For collisionless cold dark matter
particles the resulting velocity impulse is \cite{wake}, $v_{impls} 
\simeq 4\pi G\mu v_{st} \gamma_{st}$ (where $v_{st}$ is the transverse 
velocity of the string).  This leads to a wedge like region of overdensity, 
with the wedge angle being of order of the deficit angle, i.e. $8\pi G\mu$
($\sim 10^{-5}$ for GUT strings). The density fluctuation in the wake is 
of order one. Subsequent growth of this wake by gravitational instability 
has been analyzed in great detail in literature \cite{wake}. 

The structure of this wake is easy to see for collisionless particles 
(whether non-relativistic, or relativistic). Each particle trajectory 
passing by the string bends by an angle of order $4\pi G\mu$ towards the 
string. In the string rest frame, take the string to be at the origin, 
aligned along the z axis, such that the 
particles are moving along the $-x$ axis. Then it is easy to see that 
particles coming from positive $x$ axis in the upper/lower half plane will  
all be above/below the line making an angle $\mp 4\pi G\mu$ from the 
negative $x$ axis. This implies that the particles will overlap in the 
wedge of angle $8\pi G\mu$ behind the string leading to a wake
with density twice of the background density. One thus 
expects a wake with half angle $\theta_w$ and an overdensity
$\delta \rho/\rho$ where \cite{wake},

\begin{equation}
\theta_w \sim 4\pi G\mu, \qquad {\delta \rho \over \rho} \sim 1 .
\end{equation}

 However, our interest is in the density fluctuations generated by 
the strings at the quark-hadron transition in the early universe.
At that stage, it is not proper to take the matter as consisting
of collisionless particles. A suitable description of matter at that
stage is in terms of a relativistic fluid which we will take to be
an ideal fluid consisting of elementary particles in the QGP phase, 
and consisting of hadrons in the hadronic phase.
Generation of density fluctuations due to a cosmic string moving 
through a relativistic fluid has been analyzed in the literature
\cite{shk1,shk2,shk3}. The study in ref.\cite{shk1} focussed on the
properties of shock formed due to supersonic motion of the string 
through the fluid. In the weak shock approximation, one finds a 
wake of overdensity behind the string. In this treatment one can
not get very strong shocks with large overdensities. In refs.
\cite{shk2,shk3}, a general relativistic treatment of the shock
was given which is also applicable for ultra-relativistic string
velocities. The treatment in ref.\cite{shk3} is more complete
in the sense that the equations of motion of a relativistic fluid 
are solved in the string space-time (Eq.(1)), and both subsonic 
and supersonic flows are analyzed. One finds that for  
supersonic flow, a shock develops behind the string, just as 
in the study of ref. \cite{shk1,shk2}. In the treatment of 
ref.\cite{shk3} one recovers the
usual wake structure of overdensity (with the wake angle being
of order $G\mu$) as the string approaches ultra-relativistic
velocities. Also the overdensity becomes of order one in this
regime. For smaller string velocities, the results of \cite{shk1}
and \cite{shk2,shk3} are roughly in agreement. 

In the following we will reproduce some of the relevant results 
from these papers. It is not expected that the string will move
with ultra-relativistic velocities in the early universe. Various 
simulations have shown that rms velocity of string segments is 
about 0.6 \cite{vstr} for which the shock will be weak. 
In this situation, any of the approaches of refs.\cite{shk1,shk2,shk3} 
can be followed. We will briefly discuss the properties of shock,
along the lines of discussion in ref.\cite{shk1}. We will also
reproduce some relevant results from ref.\cite{shk3}. These two
sets of results will span the range of various parameters of interest 
for us. 

  The basic idea in the study of ref.\cite{shk1} 
is that the situation of a string moving through the 
fluid is the same as the situation when the fluid flows
past a wedge with the wedge angle being the 
same as the deficit angle associated with the cosmic
string. If the string velocity exceeds the sound speed in the fluid
then one gets shock waves. The properties of the shock
can be determined following the standard treatment of shocks
\cite{ll}.

   Fig.1 shows the two dimensional spatial slice, with the string
being perpendicular to the plane and located at the point marked
as O. Direction of fluid flow is shown by arrows. Dark shaded region
denotes the region of the shock with overdensity. The 
fluid region in the front of the shock has background energy
density. The wedge with white interior denotes the deficit space, with
the upper edge identified with the lower edge. Due to reflection symmetry 
of the problem, and due to the identification of upper and lower wedge 
boundaries, it is easy to see \cite{shk1} that the direction of flow must 
be parallel to the upper and lower wedge boundaries respectively
(forming the deficit region).
The situation is thus indistinguishable from the situation of fluid
flow past a wedge with wedge angle being the same as the deficit angle,
i.e. $8\pi G\mu$. We will denote the fluid variables in front of the
shock and in the back of the shock by using subscripts 1 and 2 
respectively. 

\begin{figure}[h]
\begin{center}
\leavevmode
\epsfysize=5truecm \vbox{\epsfbox{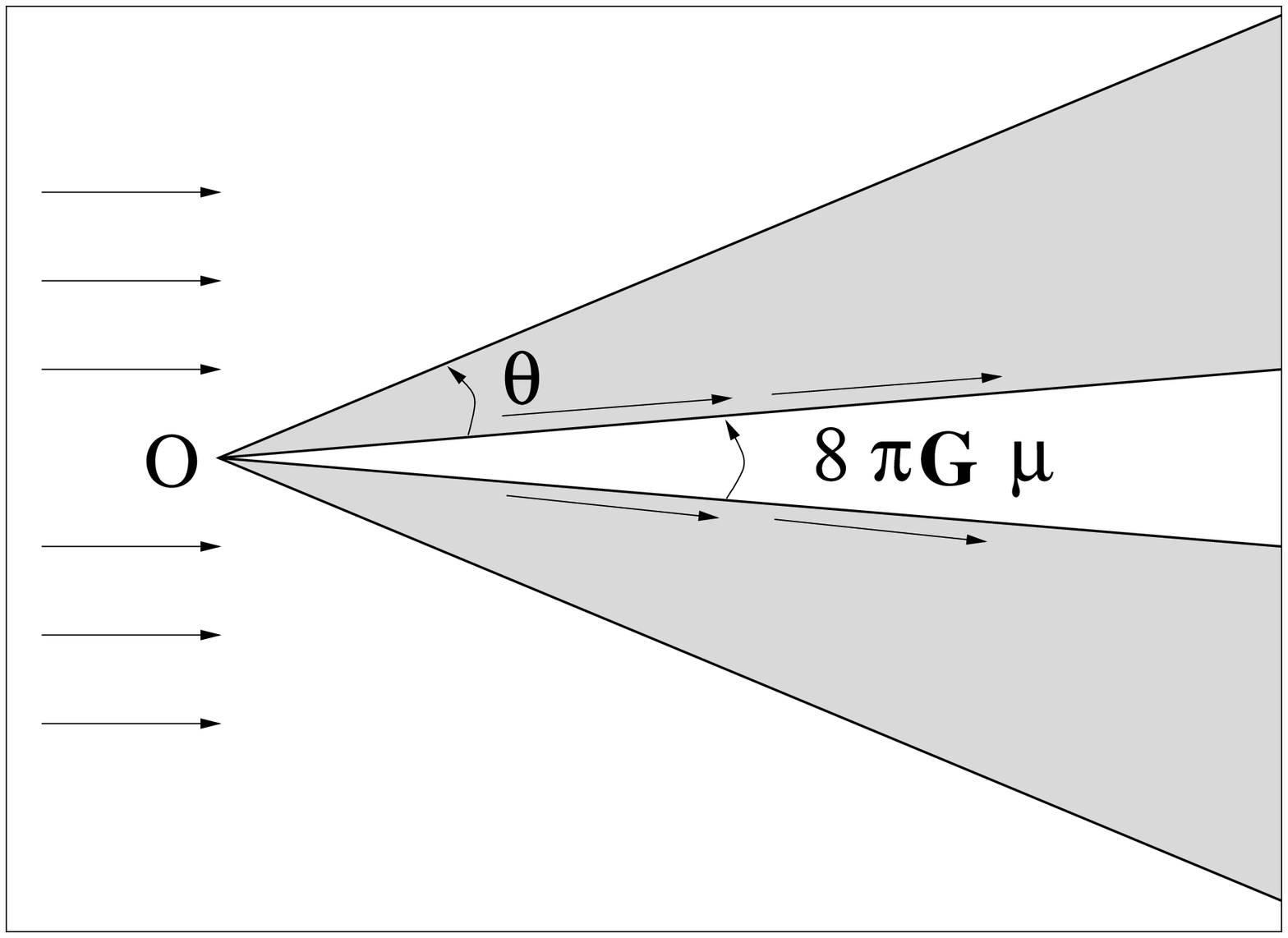}}
\end{center}
\vskip -0.3cm
\caption{}{Fluid flow past the cosmic string. Dark shaded region
denotes the region of the shock with overdensity. The
fluid region in the front of the shock has background energy
density. The wedge with white interior denotes the deficit space, with
the upper edge identified with the lower edge. Arrows denote the 
direction of fluid flow.}
\label{Fig.1}
\end{figure}

The energy momentum tensor of the fluid is taken to be of
the perfect fluid form:

\begin{equation}
T^{\mu\nu} = w u^\mu u^\nu + p \eta^{\mu \nu} .
\end{equation}

Here $w = \rho + p$,  $\rho$ and $p$ are
energy density and pressure respectively. $\mu, \nu = 0,1,2,3$.
Note that we use the flat metric $\eta^{\mu \nu}$ here as the
problem of conical space-time is now mapped to that of fluid flow
past a wedge in the Minkowski space-time.
  
 Certain continuity conditions must be satisfied at the surface
of the shock. These are \cite{ll},

\begin{equation}
T^{0n}_1 = T^{0n}_2, ~ \qquad ~ T^{nn}_1 = T^{nn}_2 .
\end{equation} 

 Here superscript $n$ denotes the direction normal to the shock surface
and subscripts 1 and 2 denote pre-shock and post-shock regions. These
conditions give (using Eq.(3)),

\begin{eqnarray}
\gamma_{1n}^2(\rho_1 + p_1)v_{1n} = \gamma_{2n}^2(\rho_2 + p_2)v_{2n} ,\\
\gamma_{1n}^2(\rho_1v_{1n}^2 + p_1) = \gamma_{2n}^2(\rho_2v_{2n}^2 + p_2) .
\end{eqnarray}

 Here $\gamma_{1n} = 1/\sqrt{1-v_{1n}^2}$ is the relativistic $\gamma$ 
factor corresponding to velocity $v_{1n}$ (similarly for $\gamma_{2n}$).
In the rest frame of the shock surface, we have,

\begin{equation}
v_{1n} = v_f sin(\theta + 4\pi G\mu), \qquad v_{2n} = v_2 sin(\theta) .
\end{equation}

$v_f$ is the velocity of the fluid in the cosmic string rest frame in 
region 1. Thus, $v_f$ is the string velocity $v_{string}$ in the
background plasma rest frame. Now, in addition to the continuity conditions 
in Eqs.(5),(6), we also require continuity of the component of velocity 
parallel to the shock. That is,

\begin{equation}
v_2cos(\theta) = v_1cos(\theta + 4\pi G\mu) .
\end{equation}

 By writing the overdensity $\Delta \rho = \rho_2 - \rho_1$ and
overpressure $\Delta p = p_2 - p_1$, (and using $G\mu << 1$), one can 
determine various properties of the shock using above set of equations. 
We get,

\begin{equation}
{\delta\rho \over \rho_1} \simeq {16\pi G\mu v_f^2 \over 
3\sqrt{v_f^2 - v_s^2}}, \qquad sin\theta \simeq {v_s \over v_f} .
\end{equation}

 For the first equation above we have used $p_1 = \rho_1/3$ appropriate
for relativistic ideal fluid. For the second equation we have used 
${\Delta p \over \Delta \rho} = v_s^2$ where $v_s = 1/\sqrt{3}$ is the 
isentropic sound speed. (To show that the shock is isentropic, 
one needs to consider additional continuity condition for 
particle number, see \cite{shk1,ll}).

  Eq.(9) contains all the relevant information for the overdense
region behind the string. The regime of validity of these equations
is $\Delta \rho/\rho, \Delta p/p << 1$. For ultra-relativistic case, we use
expressions from ref.\cite{shk3}. These are expressed in terms
of fluid and sound four velocities, (in weak shock limit),

\begin{equation}
{\delta\rho \over \rho_1} \simeq {16\pi G\mu u_f^2(1+u_s^2) \over 
3u_s\sqrt{u_f^2 - u_s^2}}, \qquad sin\theta \simeq {u_s \over u_f} ,
\end{equation}

\noindent where $u_f = v_f/\sqrt{1-v_f^2}$ and $u_s = v_s/\sqrt{1-v_s^2}$. 
We see that in the non-relativistic regime Eq.(9) and Eq.(10) give roughly
similar results. However, in the ultra-relativistic regime the shock 
properties obtained from these two sets of equations are very different 
(still within the regime of validity of the two treatments). The 
parameters of relevance for us are $\theta$ and $\delta \rho/\rho$. 
In Fig.2 we have given the plots of the values of these as obtained by 
the above two sets of equations. As we are interested in temperatures 
close to $T_c$, in a close enough neighborhood of $T_c$, the speed of 
sound becomes very small \cite{csmc}. Thus,
in Fig.2 we give plots for the value of speed of sound 
$v_s = 1/\sqrt{3}$ (shown by thick curves), as well as
$v_s = 0.1$ (thin curves) \cite{csmc}.
 
\begin{figure}[h]
\begin{center}
\leavevmode
\epsfysize=3.0truecm \vbox{\epsfbox{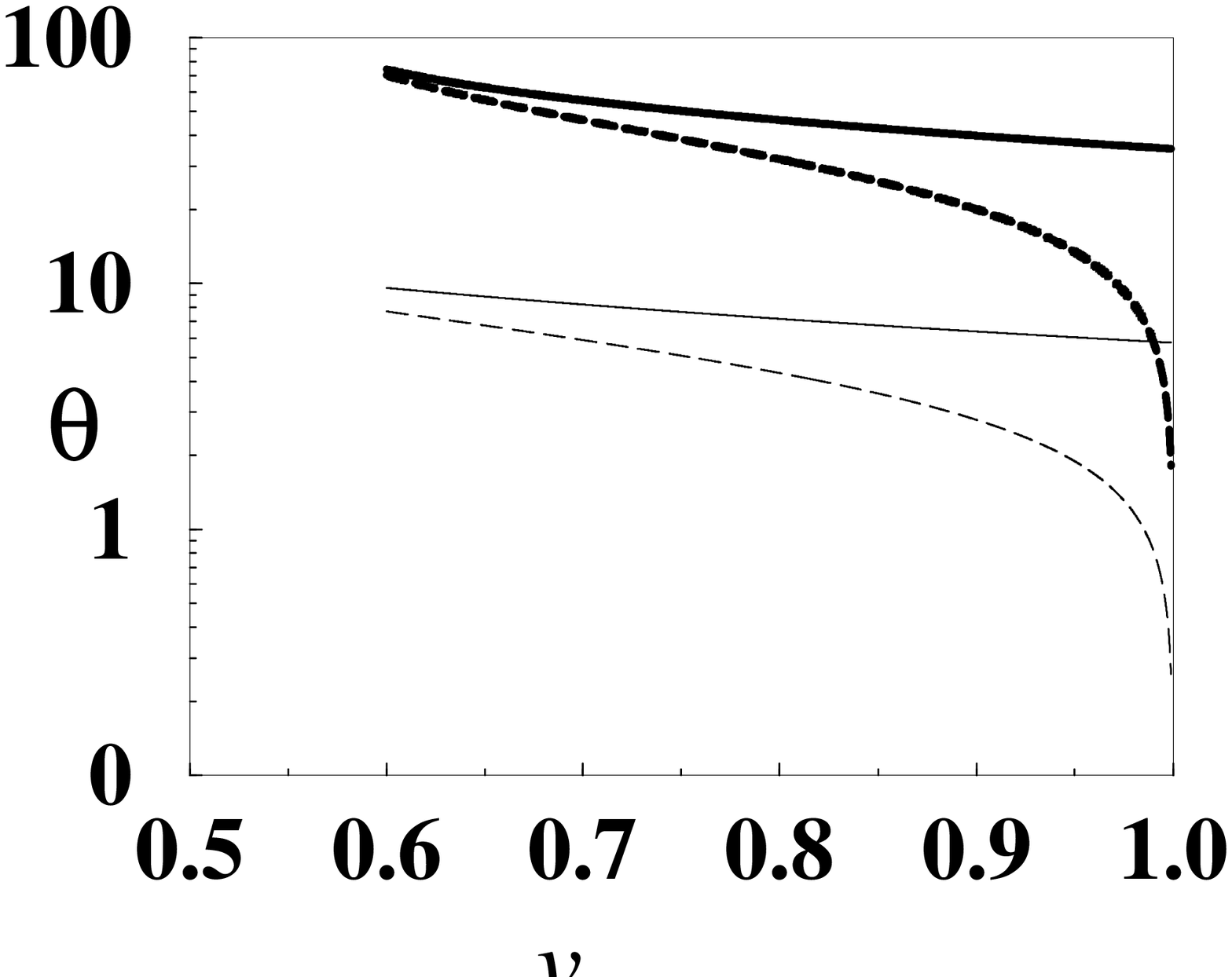}}
\epsfysize=3.0truecm \vbox{\epsfbox{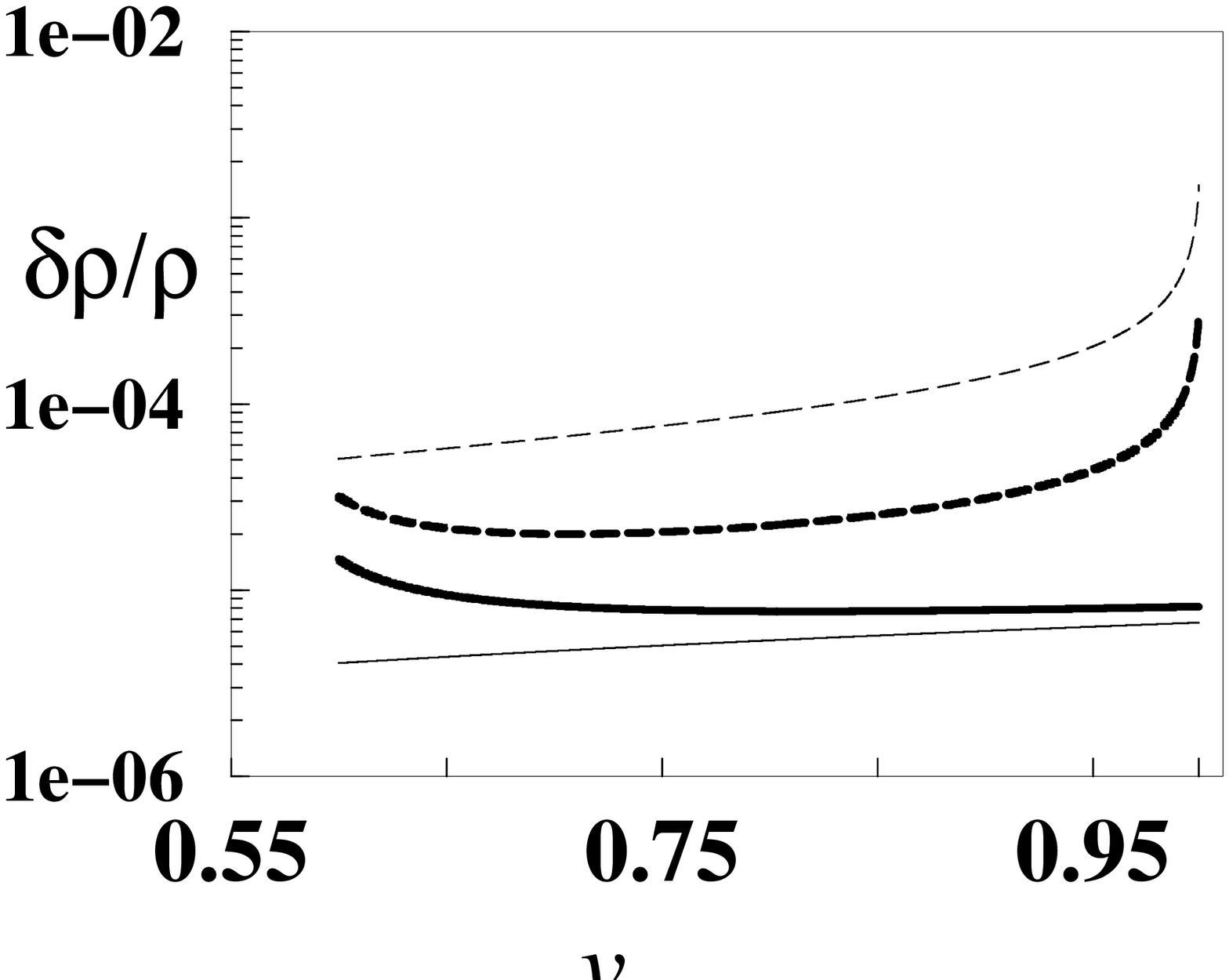}}
\end{center}
\caption{}{Plots of $\theta$ (in degrees), and $\delta \rho/\rho$, 
for a range of string velocities as obtained from Eq.(9) (solid curves) 
and Eqs.(10) (dashed curves). Thick curves denote the values 
corresponding to $v_s = 1/\sqrt{3}$, while thin curves correspond to 
$v_s = 0.1$} 
\label{Fig.2}
\end{figure}

  When string velocity is ultra-relativistic, then one can get strong 
overdensities (of order 1) and
the angle of the wake approaches the deficit angle $\simeq 8\pi G\mu$.
For string velocities in the range of 0.6 - 0.9 the predictions of
the two sets are roughly similar (within a factor of order one).
We will use a sample value corresponding to string velocity of 0.9
for which we take, 

\begin{equation}
\theta_w \simeq 20^0, \qquad  \delta \rho/\rho \simeq 3 \times 10^{-5} .
\end{equation}

 These values correspond to those obtained from Eq.10 for $v_s = 
1/\sqrt{3}$. As mentioned earlier, the overdensities expected of 
collisionless matter are of order 1, with the wake 
angle being of order of the deficit angle (as given in Eq.(2)). This is 
what one also gets from the treatment in \cite{shk3} for ultra-relativistic 
string velocities. Thus we will also consider this set of values.
With this, we are now set to get back to our problem of understanding
the effects of these density fluctuations on quark-hadron transition
in the universe. 

\section{Effect of string wakes on quark-hadron transition}

 In the conventional picture of the quark-hadron transition, the transition 
proceeds as follows \cite{wtn}. As the universe cools below the critical 
temperature $T_c$ of the transition, hadronic bubbles of size larger than a 
certain critical size can nucleate in the QGP background. These bubbles will 
then grow, coalesce, and eventually convert the QGP phase to the hadronic 
phase. The details of this transition lead to an interesting and complex 
picture \cite{wtn,csmc,fulr,kjnt,tdur,impur,inhm}. Very close to $T_c$ the 
critical size of the bubbles is too large, and their nucleation rate too 
small, to be relevant for the transition. Universe must supercool down to a 
temperature $T_{sc}$ when the nucleation rate becomes significant. The 
actual duration of supercooling depends on various parameters such as
the values of surface tension $\sigma$ and the latent heat $L$. We take
the values of these parameters as in ref.\cite{inhm} (motivated by lattice 
simulations \cite{lattice}), $\sigma = 0.015 T_c^3$ and $L \simeq 3 T_c^4$. 
With these values, one can estimate the amount of supercooling to be 
\cite{inhm,kjnt} (we take $T_c = 150$ MeV),

\begin{equation}
\Delta T_{sc} \equiv  1 - {T_{sc} \over T_c} \simeq 10^{-4} .
\end{equation}

 As the universe cools below $T_{sc}$, bubbles keep getting nucleated and
keep expanding. This nucleation process is very rapid and lasts only
for a temperature range of $\Delta T_n \simeq 10^{-6}$, for a time duration 
of order $\Delta t_n \simeq 10^{-5} t_H$ ($t_H$ is the Hubble time) 
\cite{tdur,inhm}. The latent heat released in the process of bubble expansion 
re-heats the universe. Eventually, the universe is reheated enough so that no
further nucleation can take place. Further conversion of the QGP phase
to the hadronic phase happens only by the expansion of bubbles which have 
been already nucleated. Even this expansion is controlled by how the latent
heat is dissipated away from the bubble walls. Essentially, the universe
cools little bit, allowing bubbles to expand and release more latent heat.
After the phase of rapid bubble nucleation, the universe enters into the
slow combustion phase \cite{wtn}. As we will show below, the picture of this 
slow combustion phase is very different in our model. Here, in the wake region
of the cosmic string, the conversion of the QGP phase to the hadronic phase 
proceeds by moving macroscopic planar boundaries. First, let us recall 
\cite{fulr} the estimate of the duration of this slow combustion phase.

 Energy density, entropy density, and pressure ($\rho_q, s_q, p_q$) in the 
QGP phase and ($\rho_h, s_h, p_h$) in the hadronic phase are,

\begin{eqnarray}
\rho_q =  g_q a T^4 + B, ~ s_q = {4 \over 3} g_q a T^3, ~ p_q = 
{g_q \over 3} a T^4 - B ,\\ 
\rho_h =  g_h a T^4, ~ s_h = {4 \over 3} g_h a T^3, ~ p_h = 
{g_h \over 3} a T^4 .
\end{eqnarray}

Here $g_q \simeq 51$ and $g_h \simeq 17$ are the degrees of freedom
relevant for the two phases respectively (taking two massless quark flavors
in the QGP phase, and counting other light particles) \cite{fulr}.
At the transition temperature, we must have $p_q = p_h$ which relates 
$T_c$ and the bag constant $B$, 

\begin{equation}
B = {1 \over 3} a T_c^4 (g_q - g_h) .
\end{equation}

For the transition occurring at the temperature $T \simeq T_c$, the 
latent heat is, $L = T_c (s_q - s_h) = 4B$. For the time evolution of
the scale factor $R$ of the universe, Einstein's equations give,

\begin{equation}
{{\dot R(t)} \over R(t)} = \sqrt{{8\pi G \rho(t) \over 3}} =
A \sqrt{{\rho \over B}} ,
\end{equation}

\noindent where $A^{-1}$ gives the typical time scale at the quark-hadron 
transition.  For $T_c = 150$ MeV, $A^{-1} \simeq 64 \mu sec$. Now, 
conservation of the energy-momentum tensor gives,

\begin{equation}
R(t)^3 {dp(t) \over dt} = {d \over dt} \{ R(t)^3 [\rho(t) + p(t)] \} .
\end{equation}

 For the mixed phase, we write the average value of energy density as, 
$\rho = f\rho_q + (1-f) \rho_h$, where $f$ is the fraction of volume 
in the QGP phase. Now, during the transition, $T$ and $p$ are
approximately constant. Thus Eq.(17) gives,

\begin{equation}
{{\dot R} \over R} = {-[{\dot f}(s_q - s_h) \over 3 [f(s_q - s_h) + s_h]} .
\end{equation}

 From Eqs.(16),(18), we can write down the evolution equation for 
the fraction $f$,

\begin{equation}
{\dot f} + {A \over (\chi - 1)} [4f + {3 \over \chi - 1}]^{1/2}
\{3f(\chi - 1) + 3\} = 0 ,
\end{equation}

\noindent where we have defined $\chi = {g_q \over g_h} \simeq 3$. From 
this equation one can estimate the total duration of the quark-hadron
transition $\Delta t_{trnsn}$, that is the time in which the fraction 
$f$ decreases from initial value 1 (QGP phase) to final value 0 
(hadronic phase). We find that $\Delta t_{trnsn} \simeq 14 \mu$ sec
\cite{fulr}. Compared to this the Hubble scale at the time when $T = T_c$
is $t_H \simeq 50 \mu$ sec.

 Let us now see how density fluctuations in the shocks created by cosmic 
strings affect the dynamics of quark-hadron transition. We first take the 
parameters of the density fluctuations as given in Eq.(11).  Density 
fluctuation $\delta \rho /\rho \simeq 3 \times 10^{-5}$ translates to a 
temperature fluctuation of order $ \Delta T_{wake} \equiv \delta T/T 
\simeq 10^{-5}$. We note that this temperature fluctuation is
larger than $\Delta T_n \simeq 10^{-6}$. This means that the nucleation
of hadronic bubbles will get completed in the QGP region outside the wake of 
the string, while no nucleation can take place in the wake region during
that stage. Since $\Delta T_{sc}$ is of similar order as $\Delta T_{wake}$, 
one can conclude that the outside region will enter a slow combustion 
phase before any bubbles can nucleate in the region of overdensity in 
the wake. For this it is required that the overdensity in the wake should 
not decrease at the time scale of $\Delta t_n \simeq 10^{-5} t_H$. The typical 
average thickness of the overdense region in the shock $d_{shk}$ will be
(for a wake extending across the horizon),

\begin{equation}
d_{shk} \simeq  sin\theta ~ r_H \simeq 5 km .
\end{equation}

 A density fluctuation of this length scale (which is smaller than the
horizon size) primarily propagates as a plane wave (though, since $d_{shk}$
is not too small compared to $r_h$, the evolution of this overdense region 
will have two modes, one propagating like plane wave and the other growing
as $R^2$ \cite{pdm}). Typical time scale for the evolution of the overdensity
in this region, $t_{shk}$, will be governed by the sound velocity $v_s$ which
becomes very small close to the transition temperature \cite{inhm,csmc}. 
Taking $v_s \simeq 0.1$ we get,  

\begin{equation}
t_{shk} \simeq {d_{shk} \over v_s} \simeq t_H .
\end{equation}

 Thus the density (and hence temperature) evolution in the shock region 
happens in a time scale which is too large compared to $\Delta t_n$. It is 
also larger (by a factor of about 4) than $\Delta t_{trnsn}$ during which
the quark phase is completely converted to the hadronic phase in the region
outside the wake. We mention
that in our picture, we consider the time when the universe has just started
going through the quark-hadron transition, and we focus on a region in 
which a wake of density fluctuation of size of order $r_H$ has been created 
by the moving string. Essentially the region of study for us is the horizon 
volume from which the string is just exiting at the time when the universe
temperature $T = T_{sc}$. The formation of most of the region of shock thus 
happens when the temperature is still large enough compared to $T_c$ so
that the speed of sound is close to the value 1/$\sqrt{3}$. However, some
portion of the wake will certainly form when the temperature is close
enough to $T_c$, that the relevant sound speed is small, say, $v_s \simeq
0.1$. At that stage the parameters of the shock will be determined by
the plots shown by thin curves in Fig.2. The extent of the overdense
region will be governed by the time scale $t_{shk}$. Thus, if $t_{shk}$
is much smaller than $t_H$, then wake will not extend across the horizon. 

  The precise time duration, $\Delta t_{lag}$, by which the process of bubble 
nucleation in the shock region lags behind that in the region outside the 
wake is given by \cite{inhm,tdur},

\begin{equation}
\Delta t_{lag} \simeq {\Delta T_{wake} t_H \over 3 v_s^2} \simeq 
10^{-5} t_H ,
\end{equation} 

\noindent with $v_s = 1/\sqrt{3}$. $\Delta t_{lag}$ will be larger if we take
$v_s = 0.1$. $\Delta t_{lag}$ is the extra time in which the temperature 
in the wake decreases to $T_{sc}$ compared to the time when the temperature 
drops to $T = T_{sc}$ in the region outside the wake. Since $\Delta t_{lag}$ 
is of same order as $\Delta t_n$, we conclude that the region outside the wake
enters the slow combustion phase before any bubbles can nucleate in the
wake region. It is then reasonable to conclude that the latent heat released
in the region outside the wake will suppress any bubble nucleation in the 
wake especially near the boundaries of the wake. If the heat propagates to 
the interior of the wake, then the bubble suppression may extend to the 
interior of the wake also, implying that  there will simply be no
bubbles in the entire wake region. In that case, hadronic bubbles which
have been nucleated outside the wake will all coalesce and convert the
entire outside region to the hadronic phase (with occasional QGP localized
regions embedded in it). This hadronic region will be separated from
the QGP region inside the wake by the interfaces at the boundaries of
the wake, as shown in Fig.3a. 

 Further completion of the phase transition
will happen when these interfaces move inward from the wake boundaries. 
Clearly, these moving, macroscopic interfaces will trap most of the 
baryon number in the entire region of the wake (and some neighborhood) 
towards the inner region of the wake.  Finally the interfaces will merge, 
completing the phase transition, and leading to a sheet of very large 
baryon number density, extending across the horizon. Actual value of
baryon density in these sheets will depend on what fraction of baryon
number is trapped in the QGP phase by moving interfaces. It is possible
that a large fraction of the total baryon number in the wake region
may get trapped in these sheets, resulting in sheets of extremely high 
baryon densities \cite{bfluct}. The baryon number 
density on the sheet may vary by a factor of two, from one end of the sheet
to the other end, since the shock region has the form of a wedge (though, 
the portion of the shock region formed later may have different parameters 
as compared to the portion formed in the beginning due to different
sound speeds). 

  Second possibility is that the bubble nucleation is
not entirely suppressed near the center of the shock region. While
the region outside the wake converts to the hadronic phase, same may
happen at the center of the wake as well. In that case the hadronic
phase will spread from inside the wake at the same time when the hadronic
phase is moving in from outside the wake through the wake boundary. These two
sets of interfaces will then lead to concentration of baryon number in
two different sheet like regions, with the separation between the two 
sheets being of the order of a km or so. This type of evolution is
shown in Fig.3b. (Note, from Eq.(19) that the total time duration of the
transition still remains the same, irrespective of the fact that now
the transition is happening by moving planar boundaries, rather than by 
expansion of bubbles.) The final result is that, string induced shocks will
lead to either a single sheet of high baryon concentration spreading
across the horizon, or a pair of sheets again spreading across the horizon.

Typical separation between such structures will be governed by the number
of long strings in a given horizon, which is expected to be about
15 (from numerical simulations \cite{stntwrk}). Thus these sheets (or 
pairs of sheets) will be separated by a typical distance of order of a 
km which is far too large compared to any other relevant scale at that 
time which could dissipate such inhomogeneities \cite{bfluct}. These 
structures will then certainly survive until the time of nucleosynthesis 
and will affect the abundances of elements \cite{bfluct,nsynth}.  
We mention here that we are not considering the effect of density 
fluctuations produced by string loops. These will also lead to baryon
number inhomogeneities via the effects discussed here. However, these
structures will be on a more localized scale. It is more complicated
to calculate the effects of density fluctuations by oscillating loops
(especially when time scales are of crucial importance). Still, a
more complete investigation of the effects of cosmic strings on 
quark-hadron transition should include this contribution also. 

\begin{figure}[h]
\begin{center}
\leavevmode
\epsfysize=4truecm \vbox{\epsfbox{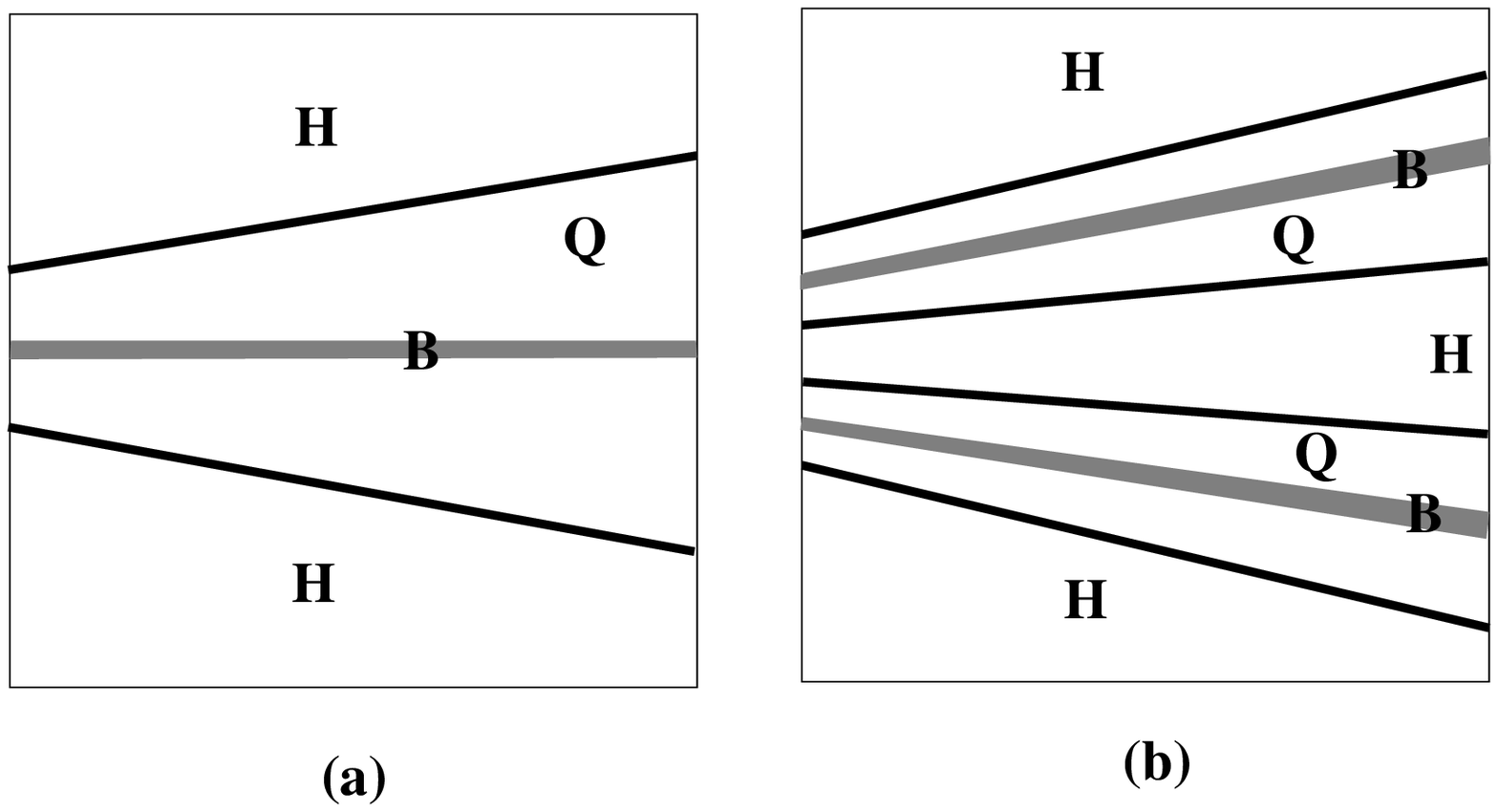}}
\end{center}
\caption{}{These figures show portions of the overdense shock region.
Q and H denote QGP and hadronic phases respectively and solid lines 
denote the interfaces separating the two phases.  (a) Situation when 
no hadronic bubbles nucleate inside the overdense wake region. 
(b) If hadronic bubbles can nucleate in the inner region of
the wake also then there will be two sets of interfaces trapping 
QGP regions in two separate sheets. Shaded regions, marked as B,
denote sheet like baryon inhomogeneities expected in the two cases.}
\label{Fig.3}
\end{figure}

 We now  discuss the situation if the shock behind the string
is strong. In that situation one expects a wake of opening angle of
about $10^{-5}$ with overdensities of order 1. These overdensities will
lead to temperature fluctuations of order 1 behind the string. This 
temperature fluctuation is far too large compared to the temperature
interval of bubble nucleation $\Delta T_n$, or even $\Delta T_{sc}$.
In such a situation, even if the transition to the hadronic phase has 
been (just) completed, as string moves through the hot hadronic
medium, the overdensity in the wake will force the temperature in
the wake region to rise above $T_c$, forcing that region to convert to
the QGP phase. The evolution of this region will be complicated
due to the tension of the interfaces at the boundaries of this region.
(Otherwise an overdensity of wavelengths much smaller than the horizon
simply propagates as plane waves). If we assume that the region primarily
cools due to the universe expansion, then a wake of QGP region extending
across the horizon will form before the interfaces close in and convert
this temporary QGP phase back to the hadronic phase. End result will again
be a sheet of large baryon density spreading across the horizon.
(In this case, the separation between the two sheets for the case of Fig.3b
will be of order few cm, as the average thickness of the wake is 
about $0.5 \times 10^{-5} \times$ 15 km $\simeq$ 8 cm. Thus, by the time
of nucleosynthesis, effectively only one sheet may survive. Also, due
to smaller wake thickness, baryon density on these sheets in this case 
will be smaller.) If the region cools at a faster rate then the 
sheets will be of smaller length.
 
\section{Conclusion}
 
 We conclude by re-emphasizing that cosmic string defects are generic
in GUT models, irrespective of their role in structure formation and it
is important to study how they can affect the dynamics of the universe.  
Dynamics of phase transitions, especially for a first order transition,
can be extremely sensitive to temperature variations. Density fluctuations,
and hence temperature fluctuations, produced due to cosmic strings,
even if they are of small magnitude, can affect the phase transition
dynamics in crucial ways. In this paper we have shown that the presence 
of cosmic strings can dramatically alter the dynamics of a first order
quark-hadron transition leading to horizon size sheets, or
pairs of sheets with large baryon number densities. These will have
strong effects on the abundances of elements as calculated during the
epoch of nucleosynthesis. This becomes important in view of the 
recent measurements of the angular power spectrum of CMBR \cite{expt}
which imply a large value of $\Omega_b$ which is not compatible 
with the calculations of conventional homogeneous nucleosynthesis. 
Our model leads to a very specific structure of baryon inhomogeneities.
These are planar structure (with large surface to volume ratio), and
are separated by a distance scale of about 1 km at the quark-hadron
transition (which translates to a distance scale of about 100 km 
when T = 1 MeV). Baryon inhomogeneities of this nature seem to be what
is required to make the results of IBBN (marginally) compatible with the
large value of $\Omega_b$ required by the recent CMBR measurements 
\cite{expt}. 

Observations of these abundances may also
be used to constrain the cosmic string parameters as well as the models
of cosmic string network evolution. Since the baryon number density
in these sheets will be very large, it will be interesting to check whether 
such sheet like regions can survive up to the present stage. Finally, we 
mention that the effects described here can lead to interesting consequences 
for electroweak (EW) phase transition. For example, just after the EW 
transition, EW symmetry may get restored in the overdense wake region 
behind the cosmic string. For a strong shock, that region may cool 
rapidly, breaking the EW symmetry spontaneously and generating baryon 
number in the process (even for a 2nd order EW phase transition). We 
will discuss this in a future work. 
 
\vskip .2in
\centerline {\bf ACKNOWLEDGEMENTS}
\vskip .1in

  We are very thankful to Sanatan Digal, Amit Kundu, Rajarshi Ray,
and Supratim Sengupta for useful comments. 


\end{multicols}
\end{document}